\documentstyle[prl,twocolumn,aps,epsf]{revtex}
\def\v#1{\mbox{\boldmath$#1$}}

\def\H{{\cal H}}
\def\HA{{\cal H}_{\rm A}}
\def\HB{{\cal H}_{\rm B}}
\def\HC{{\cal H}_{\rm C}} 
\def\JA{J_{\rm A}}
\def\JB{J_{\rm B}}
\def\JC{J_{\rm C}}

\def\B{\rm B}
\def\C{\rm C}
\def\eff{\rm eff}
\def\nn{\rm nn}
\def\nnn{\rm nnn}
\def\nnnn{\rm nnnn}

\def\mpynn{$m$-MPYNN$\cdot$BF$_4$}

\begin{document}
\draft
\title{Ground State Phase Transition in the $S=1$ Distorted Kagom\'e Heisenberg Antiferromagnets
}
\author{Kazuo Hida}
\address{Department~of~Physics,
 Faculty of Science,\\ Saitama University,  Saitama 338-8570, JAPAN}

\date{Received \today}

\maketitle

\begin{abstract}
{
The ground state phase transition in the distorted $S=1$ kagom\'e Heisenberg antiferromagnet (KHAF) is studied by means of the perturbational calculation and numerical exact diagonalization method. For strong $\sqrt{3} \times \sqrt{3}$ lattice distortion, the hexgonal singlet solid (HSS) ground state of the uniform KHAF is destroyed and new singlet state, large HSS (LHSS) state, which is globally different from the HSS state is realized. The quantum phase transition between these two singlet states is analogous to the Haldane-dimer transition in the $S=1$ antiferromagnetic Heisenberg chain. The presence of this transition supports the validity of the HSS picture of the ground state of the uniform $S=1$ KHAF.}
\end{abstract}


The kagom\'e Heisenberg antiferromagnet (KHAF) has been extensively studied theoretically and experimentally because of the interest in the interplay of the strong quantum fluctuation and the highly frustrated nature of the lattice structure. So far, most of the attempts have been focused on the ground state and low lying excitations of the uniform KHAF. In both $S=1/2$\cite{ze1,ey1,nm1,wal1}  and $S=1$\cite{kh1} cases, it is known that the ground state is a spin singlet state and the magnetic excitation has a finite energy gap. In the $S=1/2$ case, there are a number of singlet excitations below the first triplet excitation possiblly down to zero energy in the thermodynamic limit\cite{wal1}. On the other hand, the singlet excitations also have finite energy gaps in the $S=1$ case. The present author proposed the hexagonal singlet solid (HSS) picture for the ground state of $S=1$ KHAF\cite{kh1} which is analogous to the valence bond solid (VBS) picture of the ground state of the $S=1$ antiferromagnetic Heisenberg chains (AFHC)\cite{aklt}. In both HSS and VBS pictures, the $S=1$ spins are decomposed into  symmetrized pairs of two $S=1/2$ spins. In the HSS state, these $S=1/2$ spins form 6-spin singlet state around each hexagon, while in the VBS state, they form 2-spin singlet state on each bond.

As a real material, Wada and coworkers\cite{wada1,awaga1,wata1}  have investigated the magnetic behavior of \mpynn which can be regarded as the $S=1$ kagome antiferromagnet. Therefore, if the HSS picture of the ground state of $S=1$ KHAF is verified, this material is the first realistic example of the VBS-like state in 2-dimensional $S=1$ magnetic systems. In this context, it is quite important to check the validity of the HSS picture for the ground state of the $S=1$ KHAF from various points of view. In the present work, we check its validity by investigating the stability of the ground state against the lattice distortion which {\it destroys} the characteristic magnetic structure of the HSS state. 

Considering the structure of the HSS state, this state is expected to be destroyed by the lattice distortion with $\sqrt{3}\times\sqrt{3}$ structure and the phase transition to a new ground state with different magnetic structure will take place for strong enough distortion. This implies that the HSS picture of the $S=1$ KHAF is verified if the presence of such phase transition is proven. Corresponding check is well established for  $S=1$ AFHC, in which the VBS state is destabilized by strong dimerization\cite{ah,kt,kn1} and this observation elucidated that the Haldane phase is essentially different from the dimer phase in $S=1$ AFHC. Actually, we find a new ground state (large HSS state) for strongly distorted $S=1$ KHAF and a quantum phase transition between these two singlet ground states. The presence of this phase transition supports the validity of the HSS picture for the ground state of the uniform $S=1$ KHAF.

It should be also noted that the material \mpynn actually undergoes a structual transformation around 128K to the distorted phase with $\sqrt{3}\times\sqrt{3}$ structure\cite{kambe}. This is another motivation of the present study, although the strength of the distortion might not be strong enough to induce a ground state phase transition. 

\begin{figure}
\epsfxsize=60mm 
\centerline{\epsfbox{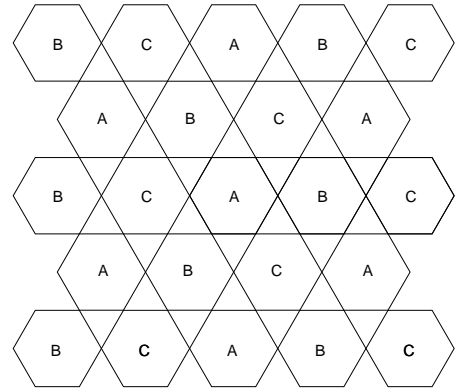}}
\vspace{5mm}
\caption{$\sqrt{3}\times\sqrt{3}$ distorted kagom\'e lattice. }
\label{fig2}
\end{figure}

Let us consider the $\sqrt{3}\times\sqrt{3}$ distorted KHAF given by,

\begin{eqnarray}
\label{ham2}
\H &=& \HA + \HB+ \HC, \\
\H_{\alpha} &=& J_{\alpha}\sum_{<i,j> \in \alpha}\v{S}_{i} \v{S}_{j}, 
\end{eqnarray} 
where $\v{S}_{i}$ is the spin operator and $\sum_{<i,j> \in \alpha}$  represent the summation over the bonds around the type-$\alpha$ $ (\alpha = \mbox{A, B} \ \mbox{or  C})$ hexgons, which are depicted in Fig. \ref{fig2}.

\begin{figure}
\centerline{\epsfxsize=60mm\epsfbox{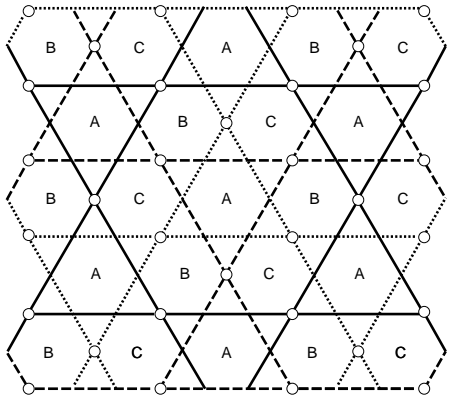}}
\vspace{5mm}
\caption{The strongly distorted KHAF. The open circles represent the 'alive' spins. The strongest effective bonds for case $\JA >> \JB=\JC$ are shown by thick lines.}
\label{kagoeff}
\end{figure}

\begin{figure}
\centerline{\epsfxsize=60mm\epsfbox{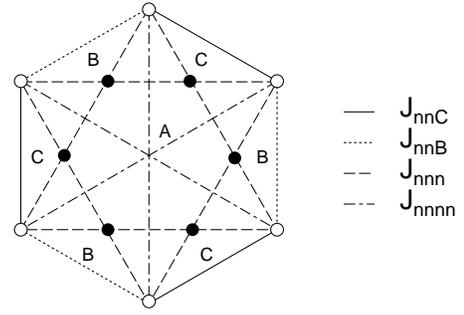}}
\vspace{5mm}
\caption{The effective exchange coupling constants between 'alive' spins (open circles).  The filled circles represent the 'dead' spins connected by the A-bonds.}\label{star}
\end{figure}

For $\JA >> \JB=\JC$, the 6 spins around each A-hexgon approximately form a 6-spin singlet state. The number of remaining spins is $N/3$.  These $N/3$ 'alive' spins form again larger kagom\'e lattice as depicted by the open circles in Fig. \ref{kagoeff}. Even in the strong distortion limit, the signs of the effective interactions between 'alive' spins are not obvious in general, because there are various paths which mediate the effective interaction. We therefore explicitly carry out the perturbation calculation up to the second order in  $\JB$ and $\JC$ to obtain the effective Hamiltonian for the 'alive' spins as,  
\begin{eqnarray}
\label{hameff}
\H_{\eff} &=& J_{\nn\B} \sum_{<\nn\B>} \v{S}_{i} \v{S}_{j} \nonumber \\
 &+& J_{\nn\C} \sum_{<\nn\C>} \v{S}_{i} \v{S}_{j} \nonumber \\
&+& J_{\nnn} \sum_{<\nnn>} \v{S}_{i} \v{S}_{j} \nonumber \\ 
&+&J_{\nnnn} \sum_{<\nnnn>} \v{S}_{i} \v{S}_{j},
\end{eqnarray} 
where  $\v{S}_{i}$ is the spin operator and the summations $\displaystyle \sum_{<\nn\B>}$,  $\displaystyle \sum_{<\nn\C>}$,  $\displaystyle \sum_{<\nnn>}$ and $\displaystyle \sum_{<\nnnn>} $ are taken over the  nearest neighbour pairs within B-hexagons, those within C hexagons, next  nearest neighbour pairs and next next  nearest neighbour pairs of 'alive' sites, respectively, as depicted in Fig. \ref{star}. 


Using the numerically obtained eigenvalues and eigenstates of the 6-spin hexagon cluster, the effective exchange couplings are calculated as,

\begin{eqnarray}
J_{\nn\B}&=&1.190573\JB\JC-0.669519\JB^2-0.519357\JC^2, \nonumber \\
J_{\nn\C}&=&1.190573\JB\JC-0.519357\JB^2-0.669519\JC^2, \nonumber \\
J_{\nnn}&=&1.112947\JB\JC-0.519357(\JB^2+\JC^2), \nonumber \\
J_{\nnnn}&=&1.035321  \JB\JC-0.519357(\JB^2+\JC^2),
\label{effj1}
\end{eqnarray}
for $S=1$ up to the second order in $\JB$ and $\JC$. Here we have set $\JA=1$.

For comparison, we have also calculated the case of $S=1/2$ as,
\begin{eqnarray}
J_{\nn\B}&=&0.388858 \JB\JC-0.251948\JB^2-0.146518\JC^2, \nonumber \\
J_{\nn\C}&=&0.388858 \JB\JC-0.146518\JB^2-0.251948\JC^2, \nonumber \\
J_{\nnn}&=&0.350555\JB\JC-0.146518(\JB^2+\JC^2), \nonumber \\
J_{\nnnn}&=&0.312253\JB\JC-0.146518(\JB^2+\JC^2).
\label{effjh}
\end{eqnarray}

If we set $\JA=1$ and $\JB=\JC=\alpha$, we have $J_{\nn\B}=J_{\nn\C}=-0.009608\alpha^2$, $J_{\nnn}=0.057519\alpha^2$, $J_{\nnnn}=0.019216\alpha^2$ for $S=1/2$ and  $J_{\nn\B}=J_{\nn\C}=0.001697\alpha^2$, $J_{\nnn}=0.074233\alpha^2$, $J_{\nnnn}=-0.003394\alpha^2$ for $S=1$.  In this case, the strongest effective interaction is the next nearest neighbour interaction. If we neglect other interactions, the whole lattice of 'alive' spins are decomposed into three equivalent sublattices of kagom\'e type which are depicted by thick solid lines, thick dotted lines and thick broken lines in Fig. \ref{kagoeff}. Therefore the ground state is expected to be the spin singlet state for $\alpha << 1$. For $S=1$, this state is the HSS state on large hexagons which are three times larger than the original kagom\'e lattice if the HSS picture is vaild for the uniform $S=1$ KHAF. We will call this state as 'large hexagonal singlet solid' (LHSS) state. It should be noted in this state that the $S=1$ spins around the A-hexagons are {\it not} decomposed into two $S=1/2$ spins but they participate the six spin singlet state as a whole. In this sense, this state is an analog of the dimer phase of the $S=1$ dimerized AFHC in which all $S=1$ spins are not decomposed into $S=1/2$ spins but paired into local singlet state as a whole. In contrast to the latter, however, the 'alive' spins in the LHSS state are again decomposed into two $S=1/2$ spins and form 6-spin singlet state on enlarged sublattices. Thus the structure of the LHSS state is globally different from  that  of the HSS state of the undistorted kagom\'e lattice.  Therefore, we may expect a ground state phase transition at a critical value of $\alpha$ from the HSS state to the LHSS state. 

\begin{figure}
\epsfxsize=50mm 
\centerline{\epsfbox{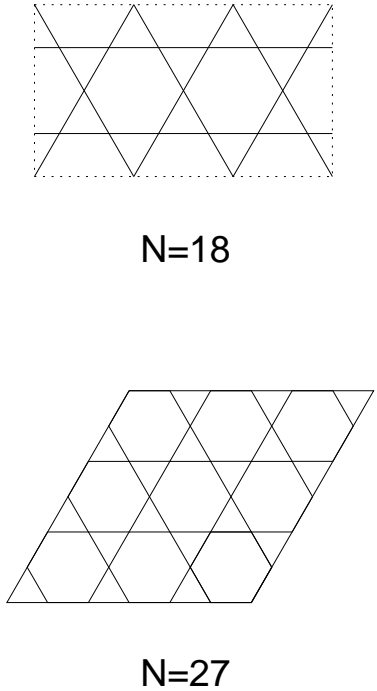}}
\vspace{5mm}
\caption{Clusters used for numerical diagonalization. }
\label{clusters}
\end{figure}
\begin{figure}
\centerline{\epsfxsize=60mm\epsfbox{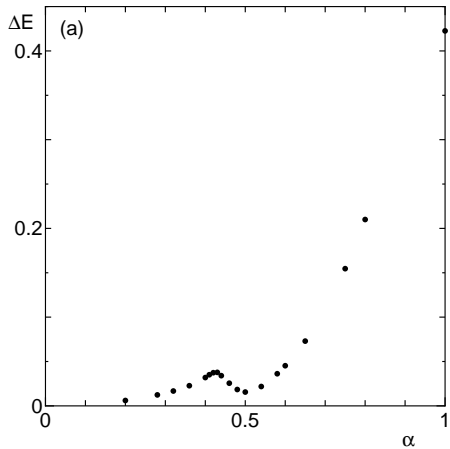}}
\centerline{\epsfxsize=60mm\epsfbox{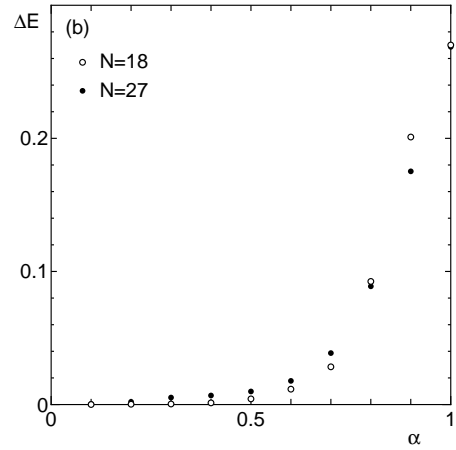}}
\vspace{5mm}
\caption{The $\alpha$-dependence of the singlet triplet gap of distorted KHAF with $\JA=1$, $\JB=\JC=\alpha$ for (a) $S=1, N=18$ and (b) $S=1/2$, $N = 18$, 27. }
\label{gapdis}
\end{figure}

To confirm the presence of phase transition between the HSS and LHSS states for $\JB=\JC=\alpha\JA$, the numerical diagonalization calculation is carried out for finite size clusters. The clusters used for the calculation are shown in Fig. \ref{clusters}. The singlet-triplet gaps are plotted in Fig. \ref{gapdis}(a) against $\alpha$ for $N=18, S=1$. It is evident that the gap has a minimum around $\alpha \simeq 0.5$ which is indicative of the phase transition in the thermodynamic limit. On the other hand, the physical picture of the ground state of the $S=1/2$ KHAF is still unclear. In this case, the singlet triplet gap is a monotonous function of $\alpha$ for $S=1/2$ as shown in Fig. \ref{gapdis}(b) for $N=18$ and 27 suggesting the absence of phase transition. Therefore, the singlet ground state of strongly distorted $S=1/2$ KHAF with $\JB=\JC$ is continuously connected with the ground state of the uniform $S=1/2$ KHAF. The presence of the HSS-LHSS phase transition for the $S=1$ case and its absence for the $S=1/2$ case support the validity of the HSS picture for the ground state of undistorted $S=1$ KHAF. 

 In general, the expressions of the effective coupling constants (\ref{effj1}) and (\ref{effjh}) show that there is a strong competition between the contributions from different interaction paths. Therefore, we may expect a rich variety of ground state phases for the distorted KHAF. For example, if we set $\JA=1$, $\JB=\alpha$ and $\JC=0$, we have $J_{\nn\B}=-0.251948\alpha^2$, $J_{\nn\C}=-0.146518\alpha^2$, $J_{\nnn}=-0.146518\alpha^2$ and $J_{\nnnn}=-0.146518\alpha$ for $S=1/2$ and $J_{\nn\B}=-0.669519\alpha^2$, $J_{\nn\C}=-0.519357\alpha^2$, $J_{\nnn}=-0.519357\alpha^2$ and $J_{\nnnn}=-0.519357\alpha^2$. Therefore all effective bonds are ferromagnetic and the ground state is ferrimagnetic with 1/3 of full magnetization. It is numerically verified that the ground state is always the ferrimagnetic state with the same value of magnetization for $\JA=1, \JB=\alpha$ and $\JC=0$ with $0 <\alpha <1$ by exact diagonalization for clusters with $N=18$. The determination of the ground state phase diagram in whole parameter plane is an interesting issue although it requires a huge computational effort.

In summary, we have found that the ground state of KHAF remains singlet even in the strong distortion limit in the case $\JA > \JB=\JC$. For $S=1$ KHAF, this singlet ground state is the LHSS state which is globally different from the HSS ground state of the undistorted $S=1$ KHAF. The HSS-LHSS phase transition is shown to take place at an intermediate value of distortion. This transition corresponds to the Haldane-dimer phase transition in the $S=1$ AFHC\cite{ah,kt,kn1}. The presence of this phase transition supports the validity of the HSS picture for the undistorted $S=1$ KHAF which is analogous to the VBS picture of the Haldane phase\cite{kh1,aklt}.

In this paper, we have discussed the ground state phase transition in detail only for  the case $\JA > \JB=\JC$, because this is the simplest case in which the ground state remains singlet even in the strong distortion limit. In general, there remains the possibility of many other kinds of ground states. The  detailed analysis of the ground state phase diagram of this model in the whole parameter plane  will be reported elsewhere. 

The numerical calculation is performed using the HITAC SR8000 at the Supercomputer Center, Institute for Solid State Physics, the University of Tokyo and the HITAC SR8000 at the Information Processing Center, Saitama University.  The numerical diagonalization program is based on the TITPACK ver.2 coded by H. Nishimori and KOBEPACK/1 coded by T. Tonegawa, M. Kaburagi and T. Nishino. This work is supported by the Grant-in-Aid for Scientific Research from the Ministry of Education, Science, Sports and Culture.

\end{document}